\documentclass[aps, pra, showpacs, twocolumn, amsfonts, amsmath, amssymb, superscriptaddress, floatfix]{revtex4}

\usepackage[T1]{fontenc}
\usepackage[latin9]{inputenc}
\usepackage{color}
\usepackage{graphicx}
\usepackage{epstopdf}
\graphicspath{pics}

\def\be{\begin{equation}}
\def\ee{\end{equation}}
\def\bea{\begin{eqnarray}} 
\def\eea{\end{eqnarray}}

\begin{document}

\title{Single-shot simulations of dynamics of quantum dark solitons}

\author{Andrzej Syrwid} 
\affiliation{
Instytut Fizyki imienia Mariana Smoluchowskiego, 
Uniwersytet Jagiello\'nski, ulica Profesora Stanis\l{}awa \L{}ojasiewicza 11 PL-30-348 Krak\'ow, Poland}
 
\author{Miros\l{}aw Brewczyk} 
\affiliation{
Wydzia\l{} Fizyki, Uniwersytet w Bia\l{}ymstoku, ul. K. Cio\l{}kowskiego 1L, 15-245 Bia\l{}ystok, Poland}
 
\author{Mariusz Gajda} 
\affiliation{
Institute of Physics of the Polish Academy of Sciences, Al. Lotnik\'ow 32/46, 02-668 Warszawa, Poland}
 
\author{Krzysztof Sacha} 
\affiliation{
Instytut Fizyki imienia Mariana Smoluchowskiego, 
Uniwersytet Jagiello\'nski, ulica Profesora Stanis\l{}awa \L{}ojasiewicza 11 PL-30-348 Krak\'ow, Poland}
\affiliation{Mark Kac Complex Systems Research Center, Uniwersytet Jagiello\'nski, ulica Profesora Stanis\l{}awa \L{}ojasiewicza 11 PL-30-348 Krak\'ow, Poland
}

\pacs{03.75.Hh, 03.75.Lm}

\begin{abstract}
Eigenstates of Bose particles with repulsive contact interactions in one-dimensional space with periodic boundary conditions can be found with the help of the Bethe ansatz. The type~II excitation spectrum identified by E. H. Lieb, reproduces the dispersion relation of dark solitons in the mean-field approach. The corresponding eigenstates possess translational symmetry which can be broken in measurements of positions of particles. We analyze emergence of single and double solitons in the course of the measurements and investigate dynamics of the system. In the weak interaction limit, the system follows the mean-field prediction for a short period of time. Long time evolution reveals many-body effects that are related to an increasing uncertainty of soliton positions. In the strong interaction regime particles behave like impenetrable bosons. Then, the probability densities in the configuration space become identical to the probabilities of non-interacting fermions but the wave-functions themselves remember the original Bose statistics. Especially, the phase flips that are key signatures of the solitons in the weak interaction limit, {can be observed} in the time evolution of the strongly interacting bosons.
\end{abstract}

\maketitle

\section{Introduction}
\label{intro}

Solitons are solutions of non-linear wave equations that maintain their shape during time evolution. They appear in different phenomena investigated in the fields of non-linear optics, ultra-cold atomic gases and in many other physical systems. Both in the optics and atomic gases bright and dark solitons have been observed experimentally \cite{KivsharOpticalSol,burger1999,denschlag2000,strecker2002,khaykovich2002,becker,Stellmer2008, Weller2008,Theocharis2010}. The former corresponds to a localized pulse propagating without a change of shape through an optical  self-focusing Kerr medium or to a localized matter-wave packet formed by attractively interacting atoms \cite{Zakharov71}. Dark solitons are related to dark holes on continuous-wave background that can be created in a self-defocusing Kerr medium or in repulsively interacting massive particles \cite{Zakharov73}. 

In the areas of non-linear optics and ultra-cold atomic gases, the non-linear wave equations are results of the approximation that assumes occupation of single modes by a large number of photons or atoms \cite{KivsharOpticalSol,pethicksmith,Kivshar1998}. These equations provide  a very accurate description of the soliton experiments \cite{Frantzeskakis10}. Evidently there are observables related to higher order correlations which cannot be obtained within a mean-field description \cite{Klaiman16}. They are however very difficult for experimental detection. Rapid development of experimental techniques in the area of ultra-cold gases opens possibility for realization and precise detection of systems with a relatively small particle number. In such a case, many-body effects that go beyond the mean-field approximation become important and experimentally relevant \cite{Theocharis2010,Lai89,Lai89a,castinleshouches,Weiss09,delande2013,corney97,corney01,martin2010b,delande2014,kronke15,
Mishmash2009_1,Mishmash2009_2,Hans2015}. 

There are examples of quantum many-body systems where solutions of the Schr\"odinger equation can be obtained analytically. In one-dimensional (1D) space, the Bethe ansatz approach turns out to be invaluable in description of bosonic or fermionic atoms and various mixtures of atomic species interacting via contact $\delta$-potentials \cite{Korepin93,Gaudin,Oelkers2006}. The celebrated Lieb-Liniger model describes bosonic particles with contact interactions in a 1D space \cite{Lieb63,Lieb63a}. In the case of periodic boundary conditions, all energy eigenstates of the system are also eigenstates of the unitary operator that translates all particles by the same distance. Thus, it is not clear which solutions of the many-body Schr\"odinger equation can be identified with the mean-field solitons because the latter break the translational symmetry. The identification is particularly difficult in the case of repulsive particle interactions where a dark soliton does not correspond to the lowest energy solution of the non-linear mean-field equation. 

Nearly 40 years ago it was conjectured that eigenstates belonging to the so-called type~II excitation spectrum of the Lieb-Liniger model were related to the mean-field dark soliton solutions due to the coincidence of the spectrum with the soliton dispersion relation \cite{kulish76,ishikawa80}. There are many publications that confirm this conjecture \cite{komineas02,jackson02,kanamoto08,kanamoto10,karpiuk12,karpiuk15,sato12,sato12a,sato16}. Especially it is shown that a proper superposition of the type~II eigenstates allows one to prepare states where dark soliton signatures are visible in the reduced one-particle density and in the behavior of the matrix element of the bosonic field operator \cite{sato12,sato12a,sato16}. Demonstration that a single type~II eigenstate reveals dark soliton character has been given recently only \cite{Syrwid2015}. The single eigenstate possesses the translational symmetry and consequently the corresponding reduced one-particle density is uniform with no soliton signature. However, one-particle density is related to density of atoms averaged over many realizations of the same experiment. For many-body systems, already a single experiment allows one to plot a particle density and such a single outcome can be very different from the average result as demonstrated in different many-body problems \cite{javanainen96,dziarmaga06,Dagnino09,Dziarmaga2010,Sacha2015,Kasevich20016}. It turns out that dark solitons can emerge in the course of measurement of particle positions \cite{dziarmaga03,dziarmaga06,delande2014}. That is, after the measurement of a single particle, the probability density for the measurement of the position of the second particle is not longer uniform because the translational symmetry is broken in the first measurement. If we continue the particle measurements, the shape of probability density for measurements of consecutive particles approaches the dark soliton profile. When $N-1$ particles is detected and only the last particle is left its wave-function matches perfectly a soliton solution of the mean-field equation \cite{Syrwid2015}.

In Ref. \cite{Syrwid2015} we have demonstrated the emergence of single and double solitons in the measurement of positions of particles prepared in type II eigenstates in the weak interaction limit. We have also analyzed results of the measurement in the case of moderate interactions. Here, we focus on dynamics of many-body states that result from translational symmetry breaking of initial symmetric eigenstates and try to answer the question if signatures of the mean-field evolution of dark solitons can be observed. Moreover, we analyze the Bose system prepared in type II eigenstates in the strong interaction regime.


The paper is organized as follows. In Sec.~\ref{liebliniger} we describe shortly the Lieb-Liniger model and present the key elements of the numerical methods that we use. In Sec.~\ref{weak} dark solitons in the weak interaction regime are considered. Section~\ref{strong} presents analysis of the system for strong interactions. Both {single and double} soliton solutions are considered. We conclude in Sec.~\ref{conclc}.

\section{Lieb-Liniger model}
\label{liebliniger}

We consider $N$ bosons with repulsive contact interactions in the 1D space with periodic boundary conditions \cite{Lieb63}. The Hamiltonian of the system in the second quantization formalism reads,
\be
H=\int_0^Ldx\left[\partial_x\hat\psi^\dagger\partial_x\hat\psi+c\hat\psi^\dagger\hat\psi^\dagger\hat\psi\hat\psi\right],
\label{h}
\ee
where $\hat\psi$ is the bosonic field operator and $L$ is the system size (the length of the 1D ring). We use units where $2m=\hbar=1$ with $m$  the particle mass. The strength of the interactions is characterized by
\be
\gamma=\frac{c}{n},
\ee
where $c>0$ is the parameter of the interaction potential and $n=\frac{N}{L}$ denotes the average density of particles. For $\gamma\ll1$ the interactions between particles are weak, for $\gamma\gg 1$ the Lieb-Liniger model describes strongly interacting impenetrable bosons \cite{Lieb63,Lieb63a}.

\subsection{Bethe ansatz solution}

The eigenvalue problem for the Hamiltonian (\ref{h}) has analytical solutions in terms of the Bethe ansatz \cite{Korepin93}. There are two branches of the excitation spectrum of the system identified by Lieb \cite{Lieb63a}. The first, called type~I, corresponds to the Bogoliubov dispersion relation in the weak interaction limit. The interpretation of the other branch (type~II) was not clear initially. Later on it was found out that the type~II branch followed the dark soliton dispersion relation in the weak interaction regime \cite{kulish76,ishikawa80}. 

Eigenstates of the Hamiltonian (\ref{h}) are determined by a set of $N$ integers (half integers) $I_j$ for odd (even) total particle number $N$ which are parameters of the Bethe equations 
\be
k_jL+2\sum_{m\ne j}^N\arctan\left(\frac{k_j-k_m}{c}\right)=2\pi I_j.
\ee
For a given {collection} $\{I\}_N$, solution of the Bethe equations results in a set of $N$ real numbers $k_j$ which are called quasi-momenta and which determine total momentum and energy of the system, i.e.
\bea
P(\{k\}_N)&=&\sum_{j=1}^Nk_j=\frac{2\pi}{L}\sum_{j=1}^NI_j, \\
E(\{k\}_N)&=&\sum_{j=1}^Nk_j^2,
\eea
respectively \cite{sato12}. The ground state of the system corresponds to the ordered set  
\be
\{I\}_N=\left\{-\frac{N-1}{2},-\frac{N-1}{2}+1,\dots,\frac{N-1}{2}\right\}.
\label{gss}
\ee
Numbers $I_j$ (integers or half-integers) are distributed symmetrically around zero,  are equally spaced by a unit distance,  and belong to the set $I_j=\pm \frac{1}{2},\pm \frac{3}{2}, \dots, \pm \frac{N-1}{2}$ for even $N$,
or $I_j=0, \pm 1, \pm 2, \dots \pm \frac{N-1}{2}$ if $N$ is odd.

The eigenstates of the system in the Lieb-Liniger model read \cite{Gaudin}
\begin{flushleft}
$\psi_{\{k\}_N}(x_1,\dots,x_N)\propto $
\end{flushleft}
\vspace{-0.3cm}
\bea
\prod\limits_{n<m}\left[\frac{\partial}{\partial{x_m}}-\frac{\partial}{\partial{x_n}}+c\;{\rm sgn}(x_m-x_n)\right]{\rm det}\left[e^{ik_jx_s}\right].
\label{ba}
\eea

It is convenient  to bring here the analogy between the studied interacting Bose system and the ideal Fermi gas in 1D. This analogy {does not have formal} character, but it provides a familiar platform to think about 
a structure of many-body eigenstates of the system. The following discussion and analogies to the Fermi gas cannot be  understood literally. 

First note, that the numbers $\{I\}_N$ are closely related to the quasi-momenta. {There is one to one correspondence between $I_j$ and $k_j$, a function relating them is monotonic (preserves ordering), i.e. if $I_i\le I_j$ then $k_i\le k_j$.}  Therefore, to simplify our discussion we will sometimes identify these two quantities addressing numbers $I_j$ as `momenta'.

If the ground state were of a form of the Slater determinant, it  could be thought of as a fully occupied `Fermi sea'.  All states with $I_{min} \le |I_i| \le I_F$ are occupied, while the others are empty.  The `Fermi momentum' here is $I_F=\frac{N-1}{2}$. The `Fermi surface' is composed of two states $\pm I_F$ of opposite momenta. The `lowest energy' state has `momentum' $I_{min}=0$ for odd $N$, while
is doubly `degenerate' for even $N$. There are two states of minimal opposite  `momenta' $|I_{min}|= \frac{1}{2}$ then.

The type~I elementary excitations correspond to the Bogoliubov  spectrum, with a phonon branch for low excitation momenta. These are particle-like excitations: the `particle from the Fermi level' is excited to the `higher momentum state'. More precisely  the `momentum' $I_F$ (or $-I_F$) is missing from the set $\{I\}_N$ and is substituted for $I_>$, where $|I_>| > I_F$.

The type~II excitation spectrum is obtained by creating a hole below the Fermi surface, substituting $|I_<| < I_F$, for the state next to the Fermi surface, $|I_f|=I_F+1$ \cite{Lieb63a,sato12,kanamoto10,Syrwid2015}. There
are two accessible `momenta' next to the Fermi surface: $I_F+1$, and the opposite one $-I_F-1$.

\subsection{Many body time correlation functions and positions' measurements}

Despite the simple form of the Bethe ansatz solution (\ref{ba}), it not easy to draw information from $\psi_{\{k\}_N}$ because number of terms in (\ref{ba}) grows dramatically with $N$. In Ref.~\cite{Syrwid2015} we have performed numerical simulations of measurements of particle positions for $N=8$ employing a sequential procedure. That is, particles have been measured one by one and at each time the conditional probability for the measurement of a next particle has been calculated \cite{javanainen96,Dziarmaga2010,Sacha2015,Kasevich20016,dziarmaga06}. The calculation of the conditional probability is possible because matrix elements $\langle\psi_{\{k^\prime\}_{N-1}}|\hat\psi|\psi_{\{k\}_N}\rangle$, so-called form factors of the field operator, can be easily obtained by means of the determinant formulas \cite{kojima97,caux07}.

It turns out that the simulations of the particle measurements can be significantly simplified with the help of Monte Carlo algorithm of Metropolis et al. \cite{Metropolis1953}, based on the Markovian walk in the configuration space. Having $|\psi_{\{k\}_N} (x_1,\dots,x_N)|^2$, we can employ the  Metropolis procedure which allows us to obtain a sequence of random sets of particles' position $\{x\}_N \equiv \{x_1,\dots,x_N\}$ from the $N$-dimension probability distribution. This procedure is equivalent to direct sampling of the $N$-paricle probability density 
$P(\{x\}_N) \equiv |\psi_{\{k\}_N} (\{x\}_N)|^2$ and also equivalent to the sequential choice of the particle positions applied in Ref.~\cite{Syrwid2015}. In the latter case, simulations become difficult in the strong interaction regime ($\gamma>1$) because the number of the form factors of the field operator needed to obtain converged results, increases significantly. The Monte Carlo method does not suffer from this problem because the form of the eigenstates (\ref{ba}) is the same regardless of the value of the interaction strength.

In this work we want to show that the Lieb-Liniger  type~II excitations occur to be solitons not only at a given instant of time, while observed,  but, what is also very important, they propagate like genuine solitons should do, 
including {double-soliton} solutions and their collisions. To this end the  two-point time correlations have to be studied. 

To observe a time evolution of the type~II soliton, first the translational symmetry of the many body state has to be broken {during} measurement of positions of $N_i$ out of $N$ particles. The soliton gets localized, the larger $N_i$, the better. Then, the system of remaining $N_r=N-N_i$ particles evolves in time till the second measurement of positions of all {remaining} particles. The question is if the second measurement reveals the soliton-like shape, shifted in the laboratory frame with respect to the initial position of solitons {in agreement with the corresponding mean-field soliton dynamics.}

To answer this question we shall probe the conditional probability density distribution
of finding $N_r$ particles at time $t$ and at positions $\{x\}_{N_r}$, provided that previously at $t=0$ another $N_i$ particles were detected in the initial state of $N$ particles prepared in a type~II exited Lieb-Linger state.

We implement the above scheme as follows. First we  choose positions $\{x_0\}_{N_i}$ at $t=0$
as any $N_i$ values belonging to  the set $\{x_0\}_N$ maximizing the $N$-body probability distribution $P(\{x_0\}_N)$ of the given state. Alternatively, we could select them randomly.  The values of these $N_i$ positions are assumed
to be set in the course of the first measurement.  The detection leads to a reduction of the system wave function. The state of remaining $N_r=N-N_i$ particles is given by the wave function 
\be
\label{phi0}
\phi(\{x\}_{N_r}) \propto \psi_{ \{k\}_N}(\{x_0\}_{N_i},\{x\}_{N_r}),
\ee
which determines a conditional probability density $P(\{x\}_{N_r}|\{x_0\}_{N_i})$ of  remaining $N_r$ particles'
positions  at the initial time.  To get the time correlations between particles distribution at the initial time $t=0$, and at the final time $t$, we have to evolve  the initial reduced $N_r$-body state, $\phi(\{x\}_{N_r})$. Let us note, that    it is not the eigenstate of the Lieb-Linger Hamiltonian anymore. Expansion of $\phi(\{x\}_{N_r})$ in the basis of the eigenstates of the $(N_r)$-particle system allows us to get time evolution
\be
\phi(\{x\}_{N_r},t)=\sum_{ \{k\}_{N_r}}e^{-it E(\{k\}_{N_r})}\; C_{\{k\}_{N_r}} \psi_{\{k\}_{N_r}}(\{x\}_{N_r}),
\label{expan}
\ee
as well as the two-time conditional probability distribution $P\left(\{x\}_{N_r},t|\{x_0\}_{N_i},t=0\right)=|\phi(\{x\}_{N_r},t)|^2$. 
This distribution is then sampled with the help of Monte Carlo algorithm. Every sample corresponds to a measurement of positions of all $N_r$ particles at time $t$ provided that $N_i$ particles were previously found at positions $\{x_0\}_{N_i}$ at the initial moment. We expect that in the limit $N_i \rightarrow \infty$, $N \rightarrow \infty$, and $N \gg N_i$ the first measurements of $N_i$ positions $\{x_0\}_{N_i}$ disturb only infinitesimally the initial type~II eigenstate.  They simultaneously perfectly localize the soliton. The reduced state obtained after the first measurement $\phi(\{x\}_{N_r})$ should be infinitesimally close to the type ~II eigenstate of the Lieb-Liniger model, a perfect soliton again. Here we are going to study rather small systems of several particles. As we are going to show, effects of a small system will lead to some   `imperfections' as compared to mean-field solutions.

The expansion (\ref{expan}) requires calculations of $C_{\{k\}_{N_r}}=\langle \psi_{\{k\}_{N_r}}|\phi\rangle$ that is not easy because it involves multi-dimensional integration. However, it turns out that the expansion can be obtained in a simpler way. Choosing many different sets of the positions $\{x\}_{N_r}$, i.e. sampling the $(N_r)$-dimensional space, and calculating values of $\phi(\{x\}_{N_r})$ and values of eigenstates $\psi_{\{k\}_{N_r}}(\{x\}_{N_r})$ which are chosen to perform the expansion, one can obtain $C_{\{k\}_{N_r}}$ by fitting the rhs of (\ref{expan}) to the sets of values of $\phi(\{x\}_{N_r})$. Such a linear regression procedure turns out to be very efficient. 

\section{Dark solitons in the weak interaction regime}
\label{weak}

In the weak interaction limit ($\gamma\ll 1$), the ground state of the Lieb-Liniger Hamiltonian and collectively excited states can be described by the mean-field approach where all bosons are assumed to occupy the same single particle state $\psi(x)$ which is a solution of the Gross-Pitaevskii equation (GPE) \cite{pethicksmith}. In the limit of the infinite system ($L\rightarrow\infty$), the dark soliton solution of the GPE reads $\psi(x)=\sqrt{n}\tanh\left(\frac{x-x_0}{\xi}\right)$ where $x_0$ is the position of the dark soliton notch and $\xi=1/\sqrt{cn}$ is the healing length that describes the width of the notch \cite{pethicksmith}. In the case of the finite system with periodic boundary conditions, the dark soliton solution has to be modified because the hyperbolic tangent function alone is not able to fulfill the boundary conditions. If $\xi\ll L$, the dark soliton solution can be approximated by $e^{i\pi x/L}\tanh\left(\frac{x-x_0}{\xi}\right)$ which describes density profile that drops to zero at the position of the soliton but the density is not stationary because the notch moves periodically along the 1D ring. Dark solitons are stationary in a frame moving with a certain velocity $v$ with respect to the laboratory frame. That is, there are dark soliton solutions of the stationary GPE,
\be
\left[\left(-i\frac{\partial}{\partial x}-\frac{v}{2}\right)^2+2c(N-1)|\psi(x)|^2\right]\psi(x)=\mu\psi(x),
\label{gpe}
\ee
with periodic boundary conditions, given in terms of Jacobi functions \cite{carr00,kanamoto09,wu13} where $\langle\psi|\psi\rangle=1$.

For chosen values of the interaction strength $c$ and average particle density $n$, one can find intervals of $v$ where single-, double- or multiple-soliton solutions exist \cite{kanamoto09}. They are characterized by the depth of the dark soliton notch 
\be
d={\rm min}\left[|\psi(x)|^2\right]L,
\ee
and by the winding number $J$ which determines a complex phase $e^{i2\pi J}$ the wave-function $\psi(x+L)$ acquires with respect to $\psi(x)$ when the position on the ring changes by $L$ \cite{kanamoto09}.

\subsection{Single solitons}
\label{ssola}

\begin{figure}
\includegraphics[width=0.9\columnwidth]{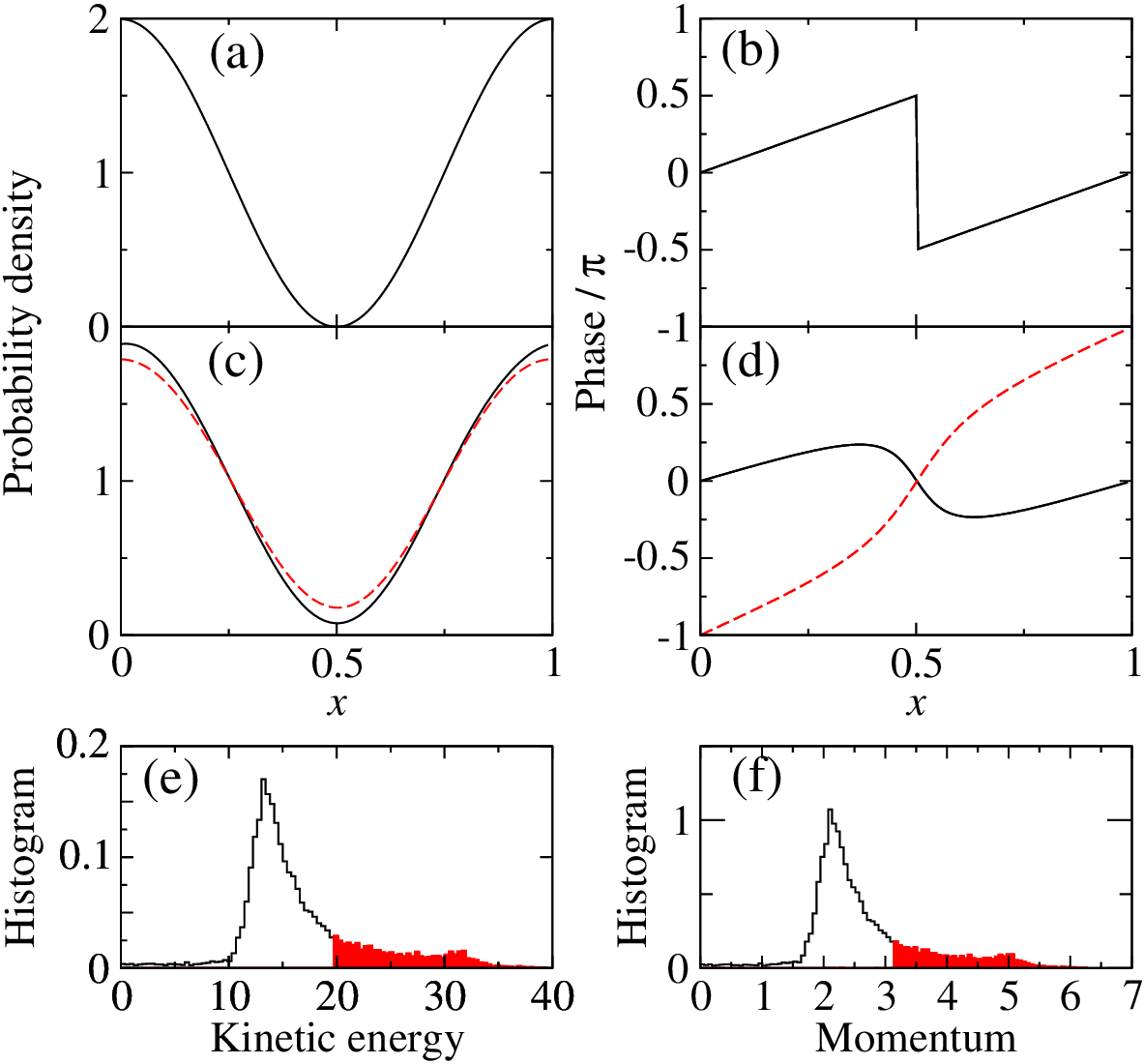}
\caption{(color online) Panels (a) and (b) show the wave-function (probability density and phase)   of the last particle (\ref{lastp}) in the measurement of positions of $N$ particles prepared initially in the one-hole excited eigenstate 
with the total momentum per particle $P/N=\pi/L$; (a) corresponds to the probability density and (b) to the phase of the wave-function. In each measurement realization, the soliton notch is localized at different position but the wave-function always matches the properly shifted mean-field solution that corresponds to the single dark soliton. The latter solution is not plotted because it would not be distinguishable from the last particle wave-function. Panels (c) and (d) show the probability density and phase, respectively, of two examples of the last particle wave-function obtained in the measurements of positions of $N$ particles prepared in the one-hole excited eigenstate with $P/N=3\pi/(4L)$. In this case there are two types of the last particle wave-functions observed in the measurement realizations. The first is related to mean-field solutions with the winding number $J=0$ [black solid lines in (c) and (d)]. The corresponding distributions of their kinetic energies and average momenta are presented in (e) and (f), respectively, see black histograms. The other wave-functions are related to $J=1$ and they are represented by red dash lines and red filled histograms in the panels (e)-(f). 
For each last particle wave-function obtained in the simulation, we can find a mean-field solution that follows it very well. Such mean-field solution is chosen so that its average momentum and $J$ are the same as those of a given last particle wave-function. The size of the 1D space $L=1$, $c=0.08$ and $N=8$.}
\label{fweak1}
\end{figure}

\begin{figure}
\includegraphics[width=1.\columnwidth]{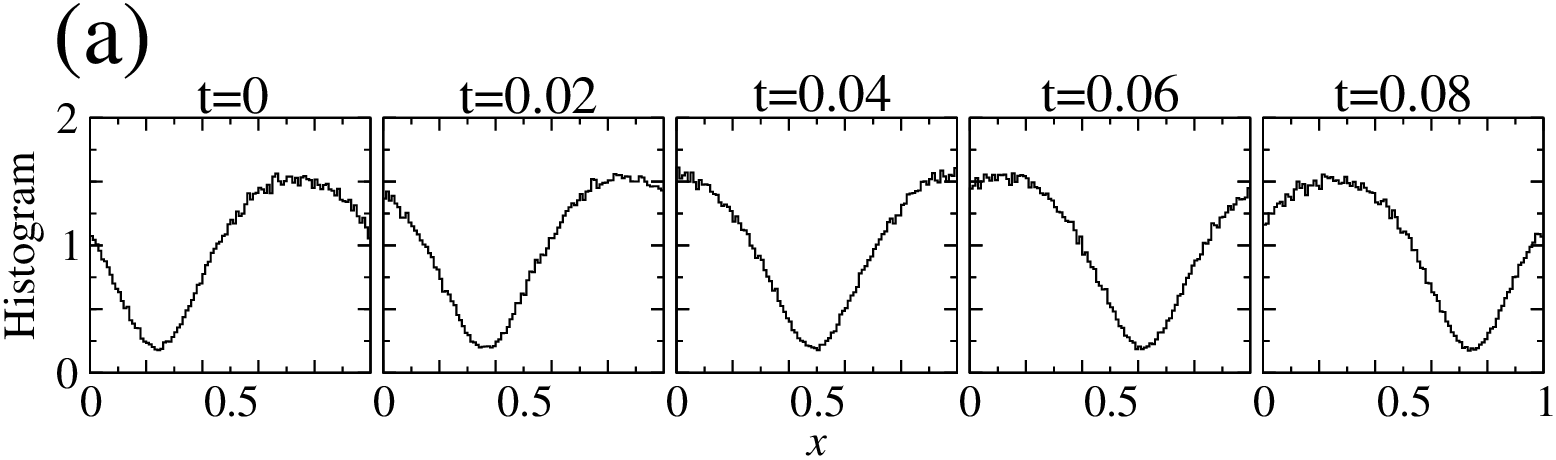}
\includegraphics[width=1.\columnwidth]{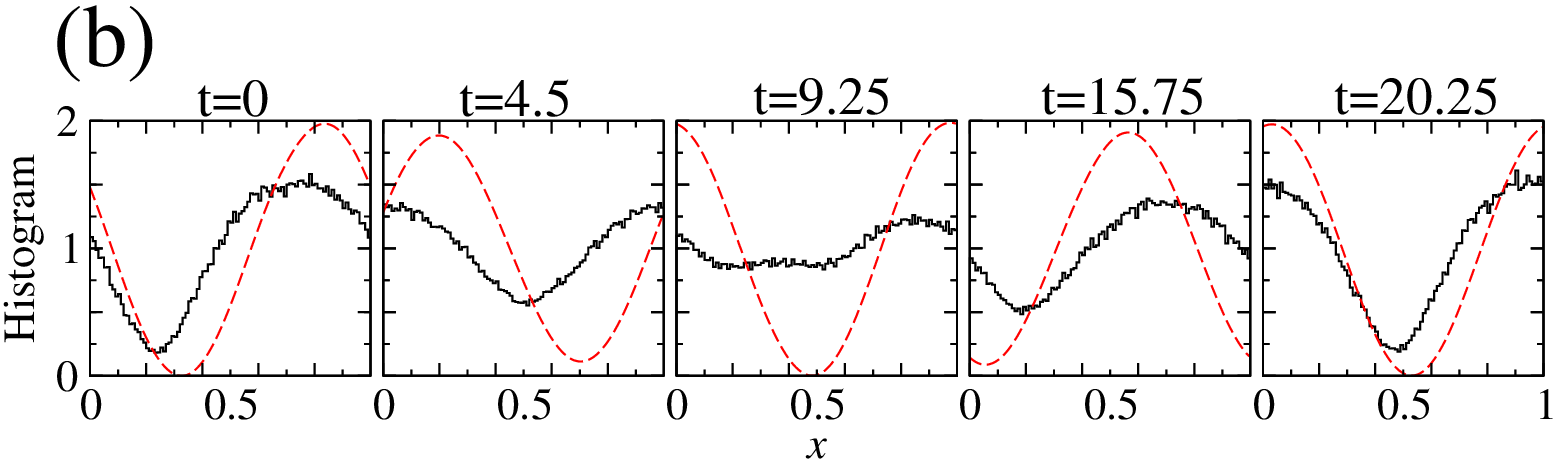}
\includegraphics[width=0.45\columnwidth]{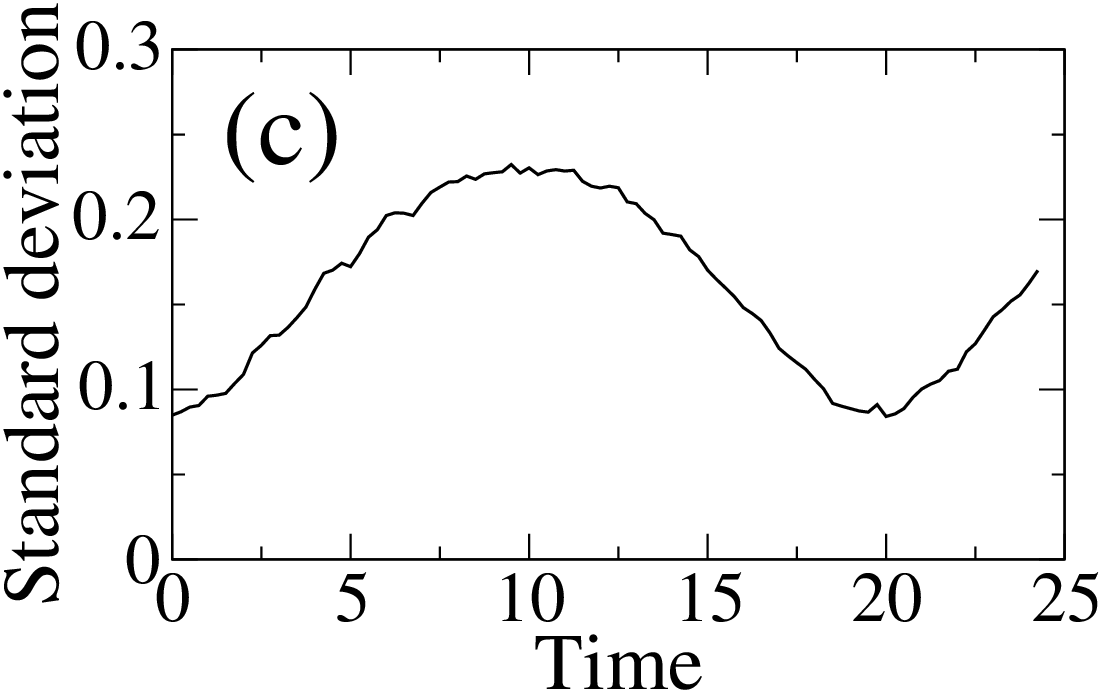}
\caption{(color online) Single soliton dynamics observed in many-particle measurement. The histogram resulting from the repeated conditional measurements of positions of $N_r=3$ particles at time $t$  proceeded by a previous detection of $N_i=N-N_r$ particles at $t=0$. 
(a) -- short time evolution.  A value of the time $t$ is  indicated in each individual panel. Initially, the $N$-particle system is prepared in the one-hole excited eigenstate with the total momentum per particle $P/N=\pi/L$. Such a many-body eigenstate is related to the mean-field dark soliton that propagates in the laboratory frame with the velocity $v=2\pi/L$ without changing its shape. Short time evolution of the histogram follows the mean-field dynamics very well. However, in the long time dynamics, presented in (b), many-body effects emerge and lead to smearing of the density notch in the histogram.
Each realization of the measurement results in a wave-function of the last particle [examples, plotted with red dash lines, are shown in (b)] that reveals dark soliton profile nearly identical to the mean-field prediction but localized at different positions. In the short time evolution the uncertainty of soliton localization around the position predicted by the mean-field description is small. However, in the course of the time evolution the uncertainty increases that is responsible for blurring of the histogram. At $t\approx 1/c$ the histogram is almost flat without clear signature of the solitons. At even longer time, we observe revival of the initial shape of the histogram.
Evolution of the uncertainty is quantitatively illustrated in (c) where the standard deviation of the notch position  of the last particle wave-functions versus time is presented.
The size of the 1D space $L=1$, $c=0.08$ and $N=8$.}
\label{fweak_dark_dyn}
\end{figure}

Let us start our comparison between the mean-field and many-body solutions with an analysis of a single-soliton which is completely dark, i.e. $d=0$. For $v=2\pi/L$ there is a single-soliton solution of the GPE (\ref{gpe}) corresponding to the average momentum $p=-i\int_0^L dx\psi^*\partial_x\psi=\pi/L$, for any value of the interaction strength $c$ \cite{kanamoto09}. The phase of the mean-field solution jumps by $-\pi$ at the soliton position. However, because the probability density drops to zero at the soliton notch, the jump of the phase can be also considered as $-\pi$ plus any multiple of $2\pi$. Thus, the winding number is meaningless in the present case.

In the many-body Lieb-Liniger model one can find a one-hole excited eigenstate (type~II eigenstate) that is characterized by the total momentum per particle $P/N=\pi/L$ and the energy per particle nearly the same as the mean-field value. The many-body eigenstate possesses the translational symmetry which is broken if we perform measurements of particle positions. The intriguing question is if the system collapses to the mean-field solution {during the measurement process.} It turns out that when we simulate many realizations of the measurement of positions of $N-1$ particles, the wave function for the last $N$-th particle matches the mean-field solution very well, see Fig.~\ref{fweak1}. The last particle wave-function is defined as follows
\be
\phi(x)\propto \psi_{\{k\}_N}(\{x_0\}_{N-1},x),
\label{lastp}
\ee
where $\psi_{\{k\}_N}$ is a chosen one-hole excited eigenstate and $\{x_0\}_{N-1}$ are fixed positions of $N-1$ particles obtained in the measurement process.
Note that in every realization of the measurement the localization of the soliton notch is random. That is, before the measurement is performed we do not know where the soliton is going to appear because the system preserves the translational symmetry \cite{Syrwid2015}.

Let us now consider a solution of the GPE that describes a gray single-soliton ($d>0$) related to the average momentum $p=3\pi/(4L)$ and to the winding number $J=0$. In the many body picture simulation of the measurement processes performed on the corresponding one-hole excited eigenstate of (\ref{h}), i.e. the state with $P/N=3\pi/(4L)$, results in a bunch of different wave-functions for the last particle. They correspond both to $J=0$ (with the probability 0.79) and to $J=1$ (with the probability 0.21). In Fig.~\ref{fweak1} we show examples of the last particle wave-functions and distributions of kinetic energies and average momenta of the last particles that we have obtained collecting results of many realizations of the measurement process. All observed wave-functions match certain mean-field single-soliton solutions.  

The choice of $|I_j|<I_F$ that is substituted for $I_F+1$ in (\ref{gss}) in a one-hole excitation determines properties of wave-functions of the last particle obtained in simulation of the measurement process. For even $N$ and for $I_j=I_{min}=\frac12$ we obtain the wave-functions that, in the mean-field approach, are related to a completely dark soliton  moving with the velocity $v=2\pi/L$, cf. Fig.~\ref{fweak1}. When the chosen $I_j$ increases and approaches $I_F-1$, the last particle wave-functions reveal gray solitons that reduce to the zero-momentum solution of the GPE (\ref{gpe}) for $N\rightarrow\infty$. On the other hand if $I_j$ decreases and reaches $-I_F$, the last particle wave-functions correspond to gray solitons that reduce to the plane wave solution of the GPE with the momentum $p=2\pi/L$ if $N\rightarrow\infty$. This is the scenario how the mean-field solitons allow us to pass smoothly between the plane wave solutions with $p=0$ and $p=2\pi/L$ \cite{kanamoto09}. Signatures of similar passing can be also observed in the many-body description. If $I_j$ is substituted for $-I_F$ {(instead of $I_F$)} we obtain analogous relation between the results of the measurements and the mean-field solutions but with $v\rightarrow -v$. 

In order to investigate dynamics of solitons we follow the procedure described in the previous section.
For a given one-hole eigenstate we choose positions of $N_i=N-3$ particles corresponding to the 
configuration of maximal probability. {In this way we break the translational symmetry of an initial eigenstate and localize} position, { with a small uncertainty,} of a soliton formed by the remaining particles. Then, we evolve the remaining $N_r=3$ particle state for a certain period of time to get the conditional two-time probability distribution. Finally we sample this distribution with the help of Monte Carlo algorithm.

In Fig.~\ref{fweak_dark_dyn} we present histograms resulting from the many-body evolution  of the previously described dark soliton that in the mean-field limit moves in the laboratory frame with the velocity $v=2\pi/L$. Short time dynamics shows that the profile of the histogram does not change and the notch, visible in the histogram, moves with the velocity predicted by the mean-field approach. However, in the long time evolution we see that the depth of the notch decreases and when $t$ is of the order of $1/c$ the notch nearly disappears. The time scale $1/c$ is much larger than the quantum speed limit time, i.e. a typical lifetime of a generic quantum state \cite{sato16,Mandelstam1945,Margolus1998,Giovannetti03}. The notch reappears again if we continue the time evolution. 

The smearing of the notch is a many-body effect which is related to an uncertainty of a soliton position \cite{corney97,corney01,martin2010b,dziarmaga03,dziarmaga06,delande2014}. 
{ The initial choice of positions of $N_i$ particles leads to localization of the soliton with a small uncertainty. That is, in different realizations of the measurement of the remaining $N_r=3$ particles, the last particle wave-functions reproduce the mean-field dark soliton profile but different profiles are localized at slightly different positions [the initial standard deviation of the soliton position is about 0.1 in Fig.~\ref{fweak_dark_dyn}(c)]. In the course of time evolution of the remaining $N_r=3$ particles, the uncertainty of the soliton position increases [see Fig.~\ref{fweak_dark_dyn}(c)] and causes smearing of the soliton notch visible in the histograms in Fig.~\ref{fweak_dark_dyn}(b).} We observe also some contributions to the histogram of particle positions coming from the last particle wave-functions corresponding to gray solitons where $d>0$. The average $d$, however, is never greater than 0.1 and these contributions can not explain the smearing of the histograms. For a longer time evolution the profile of the histograms returns to its initial shape. Such a quantum revival phenomenon can be expected in a few-body quantum system.

We are not able to perform simulations for large $N$ but one may expect that for $N\rightarrow\infty$ but with $cN=$constant, the many-body evolution follows the mean-field dynamics for time that increases linearly with $N$. 

\subsection{Double solitons}

\begin{figure}
\includegraphics[width=1.\columnwidth]{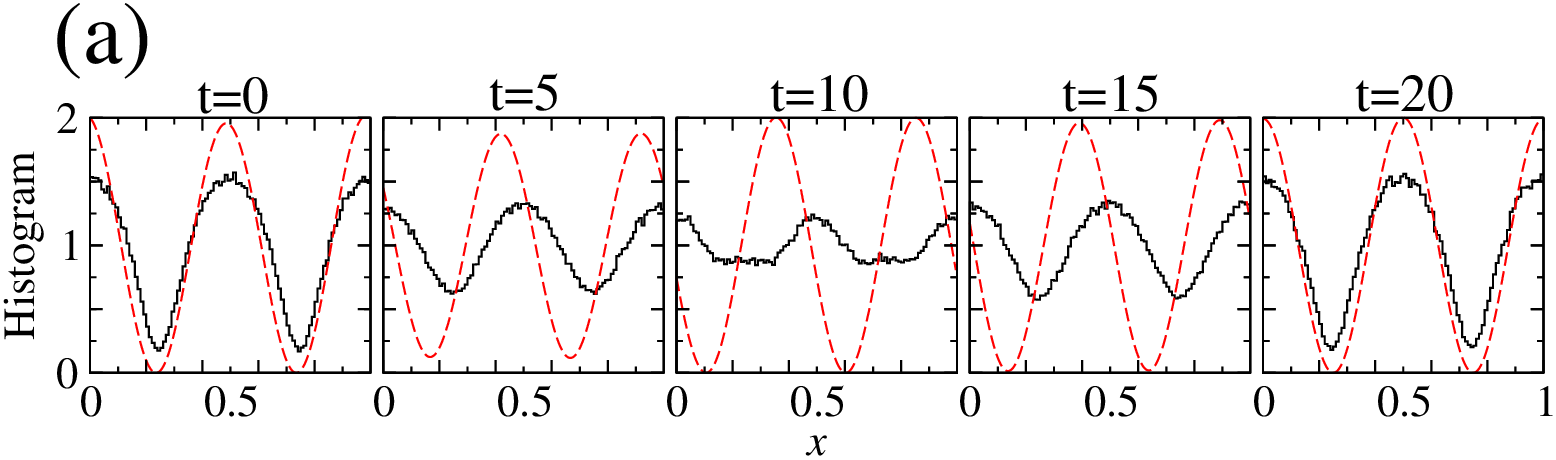}
\includegraphics[width=0.45\columnwidth]{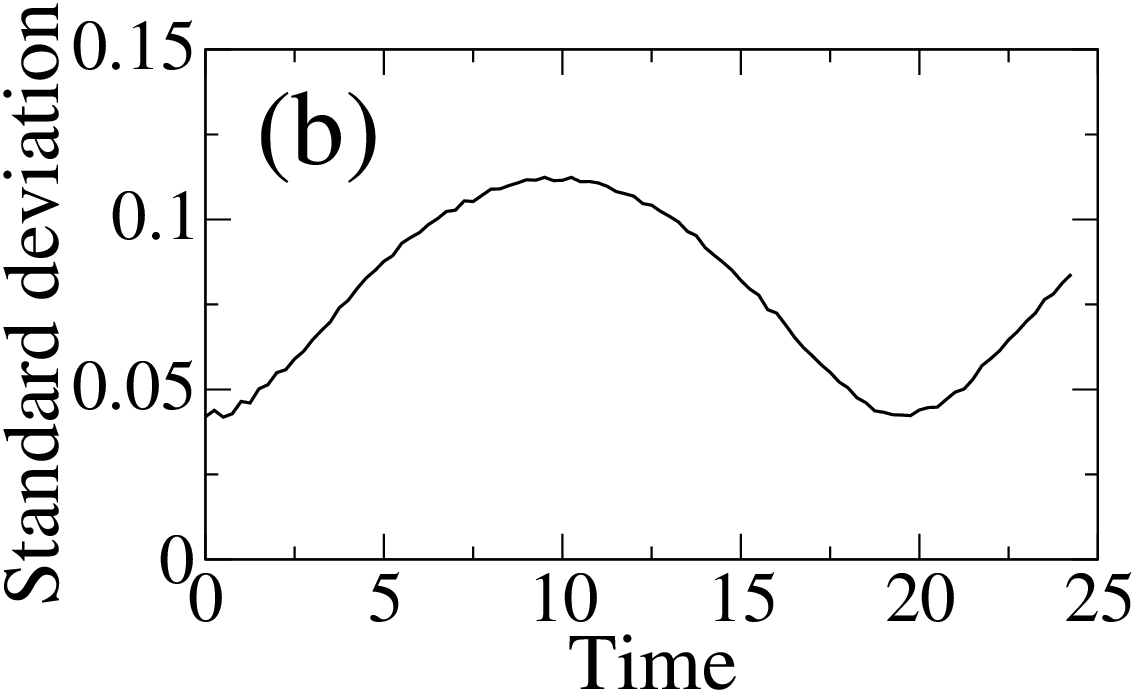}
\caption{(color online) Two soliton dynamics observed in many-particle measurement. Similar results as in Fig.~\ref{fweak_dark_dyn}(b)-\ref{fweak_dark_dyn}(c). (a) shows long time dynamics of the histogram of positions of $N_r=3$ particles obtained in many measurement realizations. Initially, the $N$-particle system is prepared in the two-hole excited eigenstate with the total momentum $P=0$, i.e., $\{-I_{min},I_{min}\} \rightarrow \{-(I_F+1),(I_F+1)\}$ in (\ref{gss}). Each realization of the measurement results in a wave-function of the last particle (examples, plotted with red dash lines, are shown in the panels) that in most of the cases matches the mean-field double dark soliton solution that is stationary in the laboratory frame ($v=0$). In each realization, the two soliton notches are always situated at the distance $L/2$ apart but they localize at different positions. At $t\approx 1/c$ the histogram is almost flat --- large uncertainty of the position of the solitons is responsible for this effect. Evolution of the uncertainty is illustrated in (b) where the standard deviation of the position of one of the notches versus time is presented. 
The size of the 1D space $L=1$, $c=0.08$ and $N=8$.}
\label{fweak2a}
\end{figure}

\begin{figure}
\includegraphics[width=1.\columnwidth]{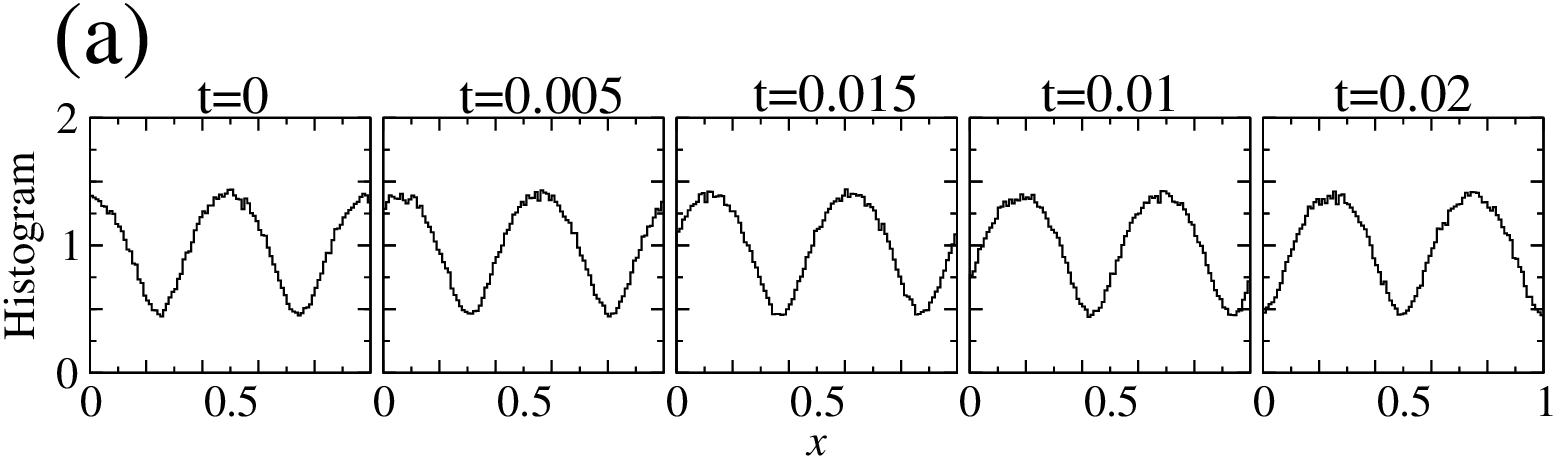}
\includegraphics[width=1.\columnwidth]{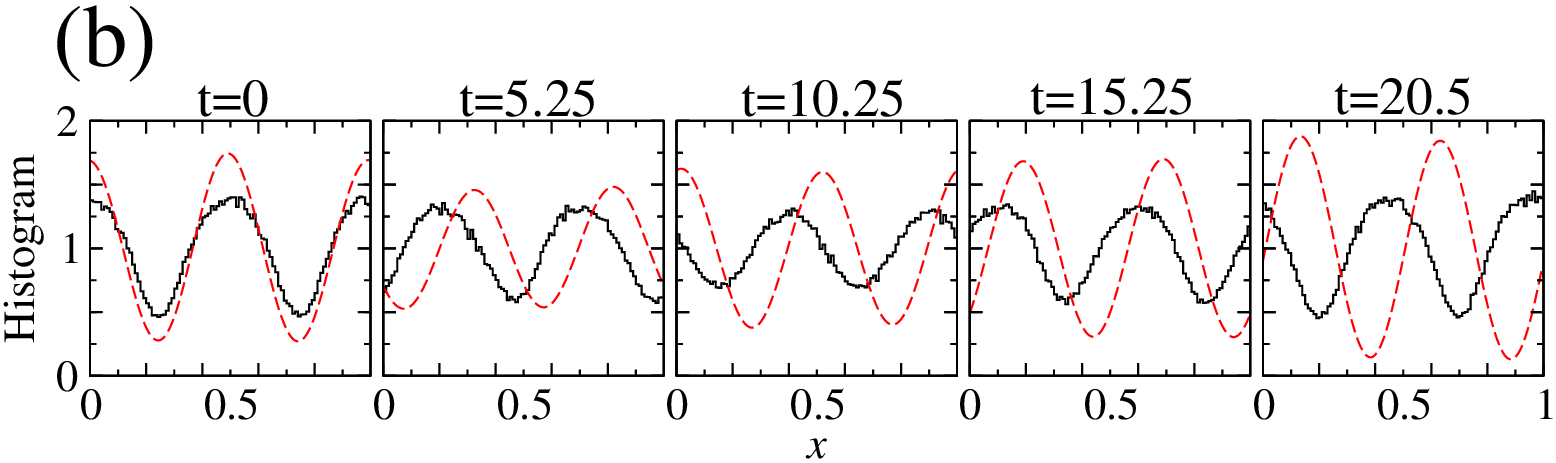}
\includegraphics[width=0.45\columnwidth]{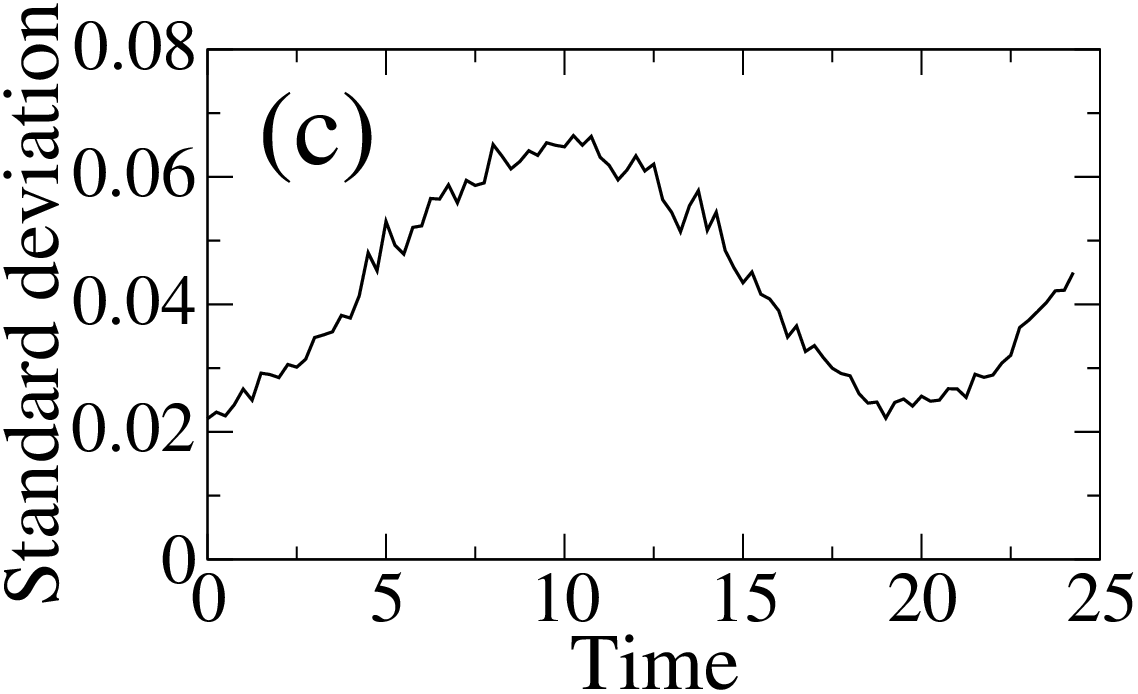}
\caption{(color online) Two soliton dynamics observed in many-particle measurement. Similar results as in Fig.~\ref{fweak_dark_dyn} but the initial $N$-body state corresponds to the two-hole excited eigenstate for $N=8$ where $\{-I_{min},-(I_{min}+1)\}\rightarrow\{(I_F+1),(I_F+2)\}$ in (\ref{gss}) that results in the total momentum per particle $P/N=3\pi/L$. The mean-field double soliton corresponding to the average momentum $p=3\pi/L$ propagates with the velocity $v\approx4\pi/L$. The notches in the histogram moves with the velocity very close to $4\pi/L$. The size of the 1D space $L=1$ and $c=0.08$.}
\label{fweak2b}
\end{figure}

\begin{figure}
\includegraphics[width=1.\columnwidth]{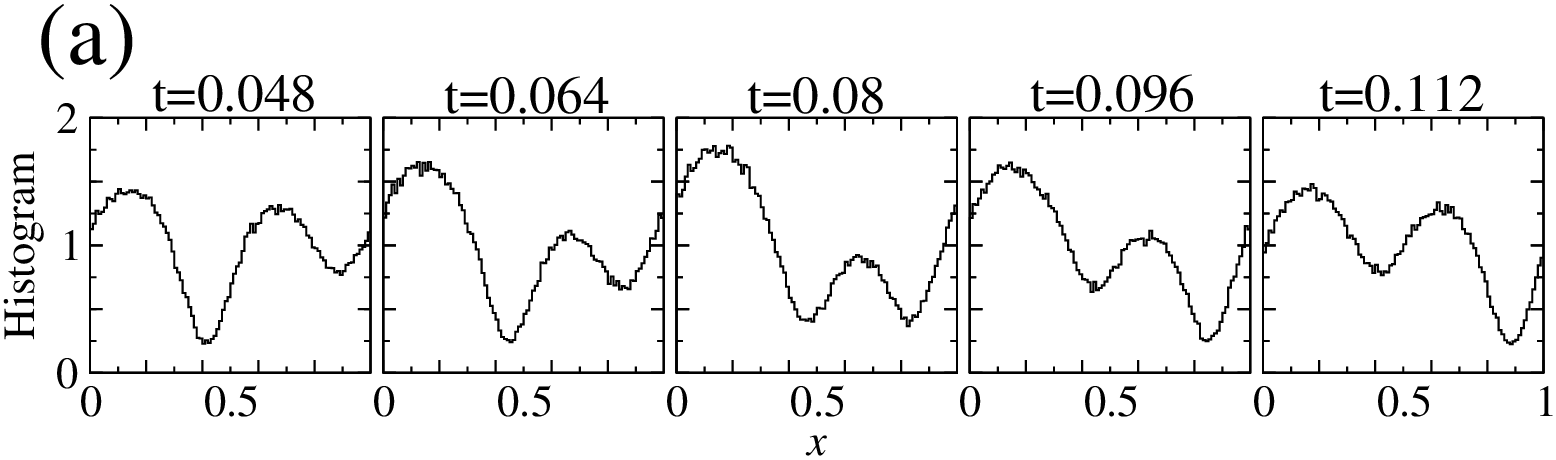}
\includegraphics[width=0.45\columnwidth]{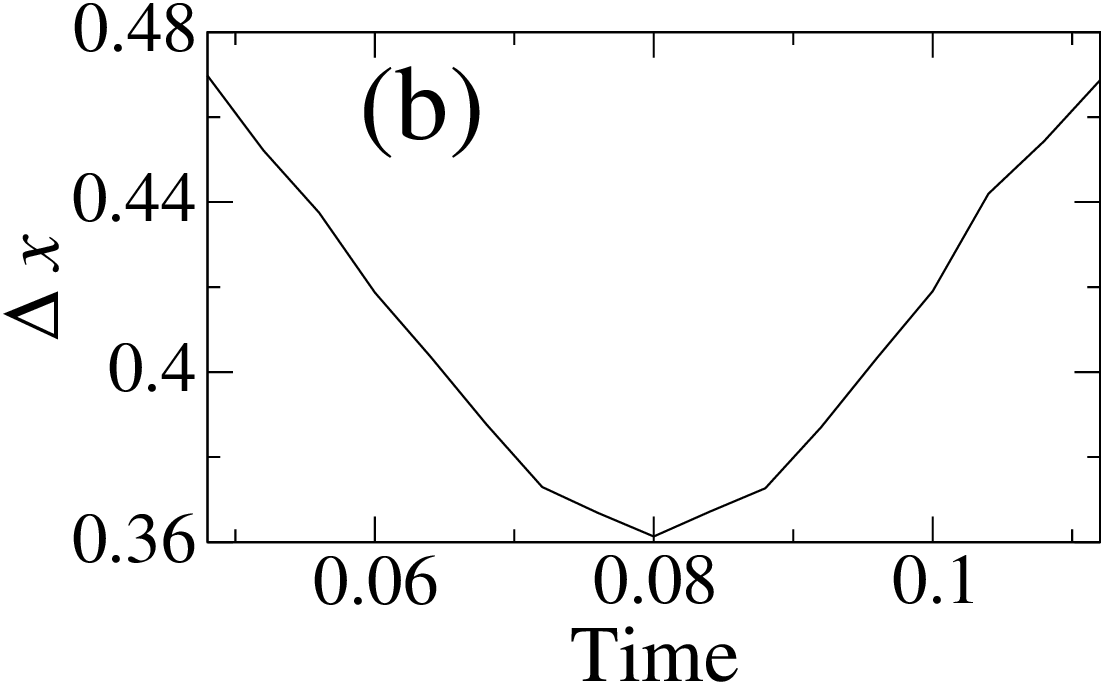}
\caption{{Signatures of collision} of two solitons observed in many particle measurement. (a): Histograms similar as in Fig.~\ref{fweak_dark_dyn}(a) and Fig.~\ref{fweak2b}(a) but the initial $N$-body state corresponds to the two-hole excited eigenstate where {$\{-(I_{min}+2),I_{min}\}\rightarrow\{-(I_F+1),(I_F+1)\}$} in (\ref{gss}). The subsequent panels show moments of time where two different minima: propagate towards each other, even their depth and exchange their positions. (b): distance $\Delta x$ between the minima, visible in (a), versus time. The size of the 1D space $L=1$, $c=0.08$ and $N=8$.}
\label{fweak2c}
\end{figure}

Double-soliton solutions of the GPE are related to two-hole excited eigenstates of the Lieb-Liniger model \cite{Syrwid2015}. That is, when we remove two values of $|I_{j_1}|<I_F$ and $|I_{j_2}|<I_F$ in the ground state sequence (\ref{gss}) and add new ones next to the Fermi surface: $-(I_F+1)$ and $(I_F+1)$ or $(I_F+1)$ and $(I_F+2)$ instead, we observe that in the measurement processes the wave-functions of the last particle match mean-field double-soliton states. 

There is a double-soliton solution of the GPE that is characterized by the average momentum $p=0$. This solution is stationary in the laboratory frame ($v=0$) and describes two completely dark solitons ($d=0$). In the many-body picture the corresponding two-hole excited eigenstate, related to $P/N=0$, can be obtained by the exchange $-I_{min}$ and $I_{min}$ for $-(I_F+1)$ and $(I_F+1)$  in the sequence (\ref{gss}) {for even $N$.} 

It is also possible to obtain a mean-field solution where two completely dark solitons move together in the laboratory frame with the velocity $v=4\pi/L$. Then, the related two-hole excited eigenstate corresponds to the substitution $\{I_{min},(I_{min}+1)\rightarrow\{(I_F+1),(I_F+2)\}$ in (\ref{gss}) {for even $N$.} When we carry out the measurements on the many-body system prepared in one of these two two-hole excited eigenstates, wave-functions of the last particle reproduce very well the corresponding mean-field solutions, obviously, in the view of the previous discussion, localized at different positions in different realizations of the measurement process.

Two-hole excitations can be chosen in many different ways. We can find two-hole excited eigenstates that reveal signatures of solutions of the GPE (\ref{gpe}) if we perform the measurement of particle positions. Equation~(\ref{gpe}) describes two (or more) identical solitons that either move together with the same velocity in the laboratory frame or do not move at all. However, one may also expect situations where solitons move with different velocities and collide. Such a process can not be described by the stationary GPE (\ref{gpe}). It is interesting if signatures of two solitons moving with different velocities can be observed in the many-body system prepared initially in a two-hole excited eigenstate.

Let us begin the analysis of soliton dynamics with the two-hole excited eigenstate corresponding to the completely dark mean-field solitons that are stationary in the laboratory frame ($v=0$). Figure~\ref{fweak2a} show how the histogram of positions of $N_r=3$ particles evolves in time if $N_i=N-3$ particles have been measured at $t=0$. Similarly to the single-soliton case, short time dynamics agrees with the mean-field predictions but in the long time evolution we observe that the notches are gradually smeared and disappear at $t\approx 1/c$. The uncertainty of the soliton positions is responsible for such a many-body phenomenon as discussed in Sec.~\ref{ssola}.

In the second example we consider a two-hole eigenstate obtained by the exchange of $-I_{min}$ and $-(I_{min}+1)$ for $(I_F+1)$ and $(I_F+2)$ in the ground state sequence (\ref{gss}) {for $N=8$} which leads to {$P/N=3\pi/L$}. The double-soliton solution of the GPE (\ref{gpe}) corresponding to the the same value of the average momentum {$p=3\pi/L$} describes two gray solitons moving together with the velocity $v\approx 4\pi/L$. Simulations show that the notches in the histogram of positions of the last $N_r=3$ particles, when the first $N_i=N-3$ particles are measured at $t=0$, move with a very similar velocity, see Fig.~\ref{fweak2b}. Wave-functions of the last particle obtained in each final measurement of the 3-particle system are different but all correspond to mean-field solitons that propagate also with $v\approx 4\pi/L$. The smearing of the soliton notches is visible for $t$ of the order of $1/c$ but it is less apparent than in the cases corresponding to the completely dark mean-field solitons.

In the third example we would like to consider a two-hole eigenstate that reveals signatures of two solitons moving towards each other with different velocities. We begin the simulation with an eigenstate where $\{-(I_{min}+2),I_{min}\}\rightarrow \{-(I_F+1),(I_F+1)\}$ in the sequence (\ref{gss}) for $N=8$, then we measure positions of $N_i=N-3$ particles ($t=0$), evolve the system of the remaining $N_r=3$ particles in time and at different $t$ perform simulations of measurements of positions of 3 particles. Analysis of the wave-functions of the last particle indicates that they may be very different. There are wave-functions that show dark and gray solitons that are clearly separated one from each other but there are also profiles where clear separation of the two solitons is not observed. The latter cases can be related to a collision of the two solitons. The average behavior of the three-particle system is represented by the time evolution of the histogram of the detected positions of three particles in Fig.~\ref{fweak2c}. We can see that two density notches (one nearly dark and the other gray) approach each other, even their depth and exchange their positions as if two sloitons collide, pass each other and propagate further.  

To sum up, two-hole eigenstates we have investigated can be divided into three classes. The first class is related to completely dark solitons in the mean-field approach. That is, when the measurement of particle positions is performed, we always observe that the last particle wave-functions match quite well dark soliton solutions with average momentum $p$ equal to the initial total momentum per particle $P/N$ of the many-body system. There are two such eigenstates. The first, obtained by the exchange $\{-I_{min},I_{min}\}\rightarrow\{-(I_F+1),(I_F+1)\}$ for even $N$, corresponds to $P/N=0$ and reveals the double-soliton mean-field solution that is stationary in the laboratory frame. The other eigenstate, related to $\{I_{min},(I_{min}+1)\}\rightarrow\{(I_F+1),(I_F+2)\}$ or to {$\{-(I_{min}+1),-I_{min}\}\rightarrow\{-(I_{F}+2),-(I_{F}+1)\}$}, where $|P/N|=2\pi/L$, reveals double-solitons that move with $|v|=4\pi/L$ in the mean-field approach.

The second class consists of eigenstates which are obtained by an exchange $\{I_j,I_{j+1}\}\rightarrow\{\pm(I_F+1),\pm (I_F+2)\}$ with $I_j\ne I_{min}$. The measurements reveal different wave-functions of the last particle that match gray double-soliton solutions of the GPE (\ref{gpe}) but corresponding to different average momentum $p$ in different measurement realizations. That is, $p$ is not necessarly equal to $P/N$ of the initially chosen many-body eigenstate.

The third class corresponds to two-hole excitations where $\{I_i,I_{j}\}\rightarrow\{\pm(I_F+1),\pm(I_F+2)\}$ or $\{I_i,I_{j}\}\rightarrow\{-(I_F+1),(I_F+1)\}$ but $j\ne i+1$. Then, the last particle wave-functions reveal double-solitons with two notches that are not identical, appear at different distances one from each other and show signatures of a collision of two solitons. Such a soliton dynamics can not be described the stationary GPE (\ref{gpe}).

\section{Strong interaction regime}
\label{strong}

\begin{figure}
\includegraphics[width=1.\columnwidth]{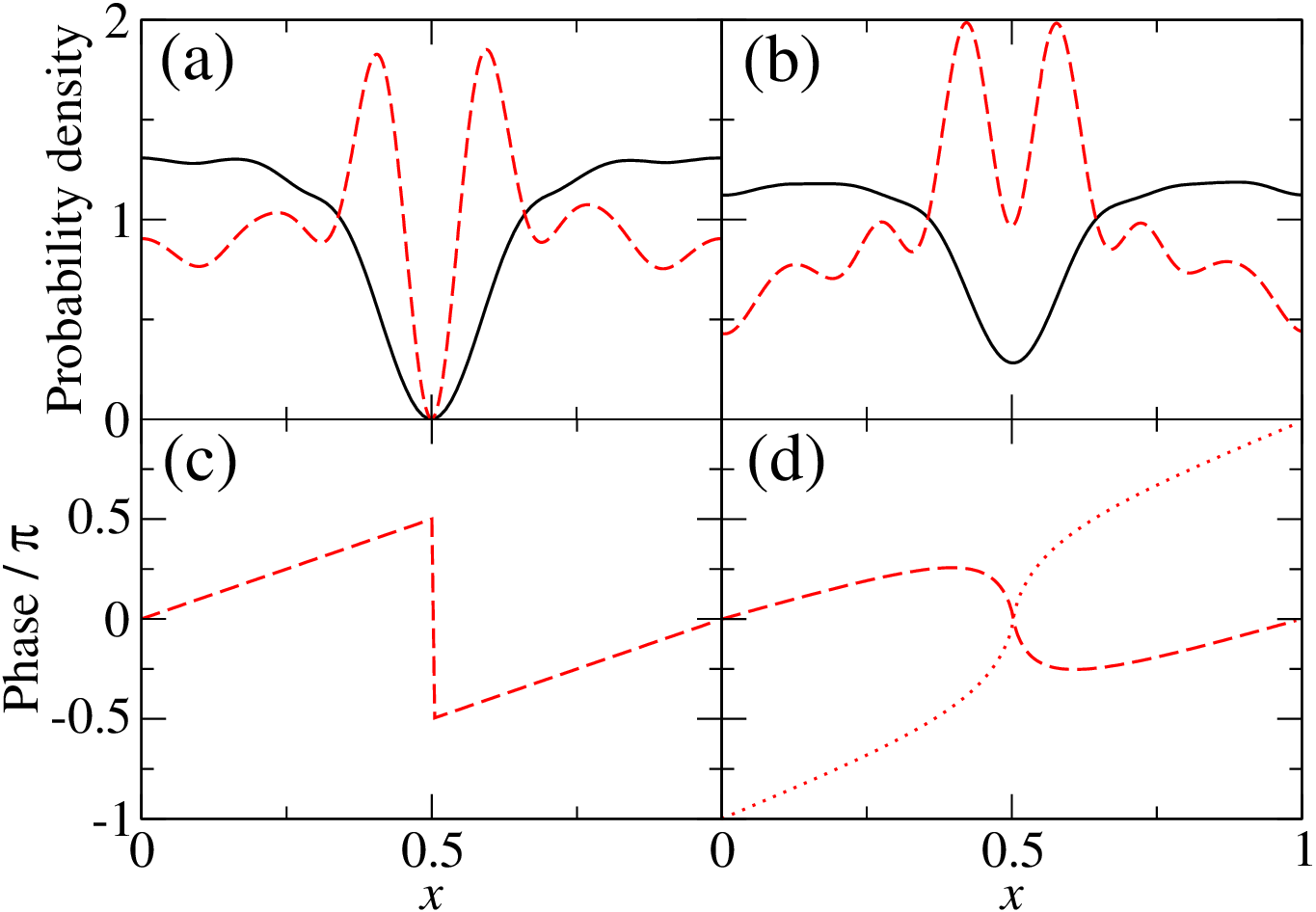}
\caption{(color online) {One-hole excitation in} a strong interaction regime. Panels (a)-(b) show average probability densities corresponding to wave-functions of the last particle obtained in simulations of the measurements of particle positions. Panels (c)-(d) present average phase of the last particle wave-functions. The phase flips appear at different positions in different realizations of the measurements but before the averaging the wave-functions are shifted so that the positions of the flips coincide with $x=L/2$. Left panels are related to the one-hole excited eigenstate obtained by the exchange $I_{min}\rightarrow(I_F+1)$ in (\ref{gss}), i.e. to $P/N=\pi/L$, while right panels to the eigenstate where $(I_{min}+1)\rightarrow(I_F+1)$ where $P/N=3\pi/(4L)$. In (a)-(b) particle interactions correspond to $\gamma=1$ (black solid lines) and $\gamma=10^3$ (red dash lines) The average phases change very little with a change of $\gamma$, therefore in (c) and (d), we show the results for $\gamma=10^3$ only. In (d) red dash line is related to the average over the wave-functions with the winding number $J=0$ and red dotted line to the average over the wave-functions with $J=1$ --- both kind of the wave-functions appear in the simulations similarly to the weak interaction case, cf. Fig.~\ref{fweak1}(d). The size of the 1D space $L=1$ and $N=8$.}
\label{fstrong1}
\end{figure}

\begin{figure}
\includegraphics[width=1.\columnwidth]{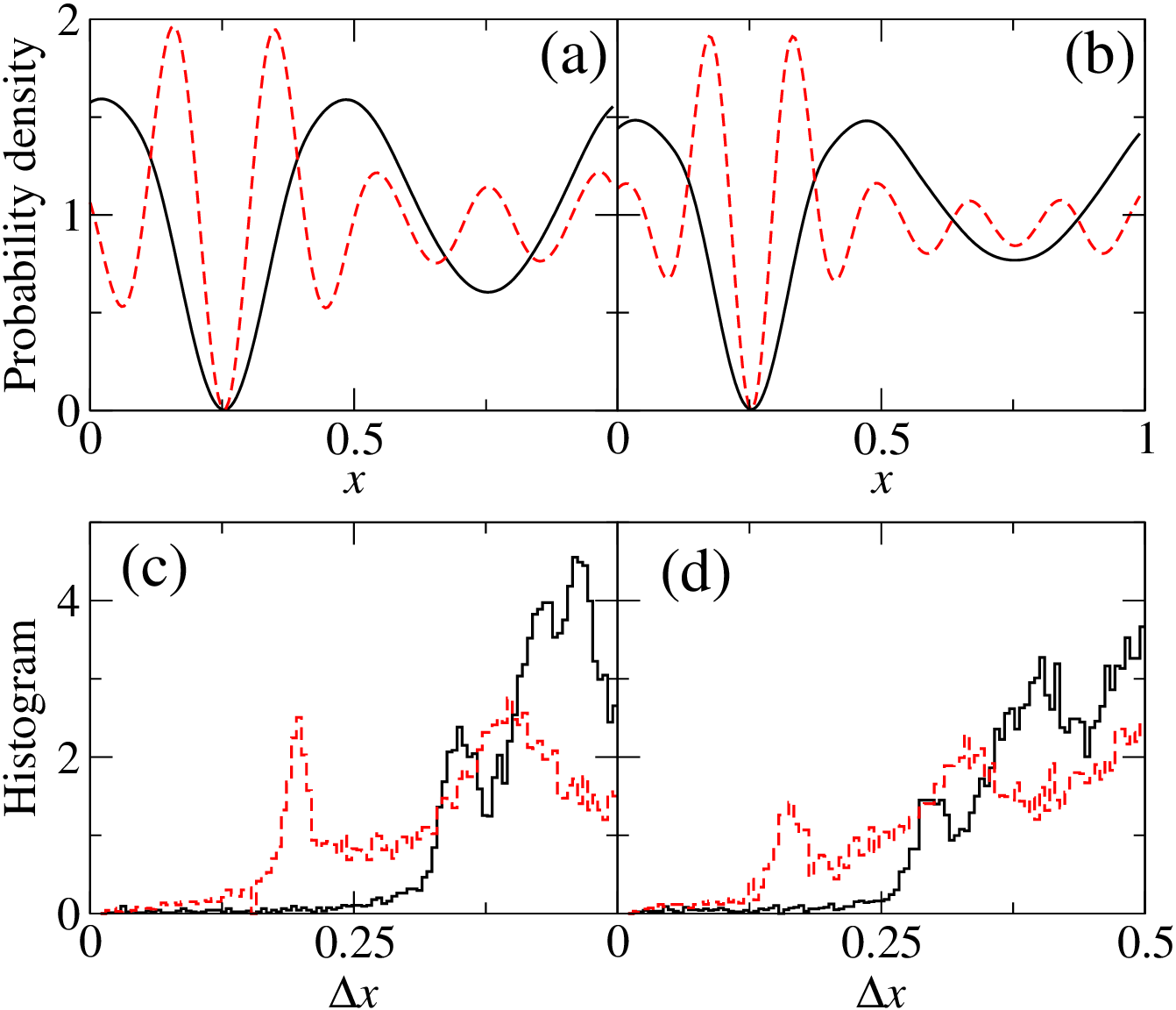}
\caption{(color online) {Two-hole state} in a strong interaction regime. Panels (a)-(b): average probability densities corresponding to wave-functions of the last particle obtained in the measurements of positions of  particles prepared initially in the two-hole excited eigenstate where $\left\{-I_{min},I_{min}\right\}\rightarrow\left\{-(I_F+1),(I_F+1)\right\}$ in (\ref{gss}). The wave-functions reveal two phase flips. Before averaging the wave-functions are shifted so that position of one of the flips coincides with $L/4$. Similarly as in {Fig.~\ref{fstrong1}(a)-\ref{fstrong1}(b)}: $\gamma=1$ (black solid lines) and $\gamma=10^3$ (red dash lines). Panels (c)-(d) show histograms of distances $\Delta x$ between the positions of the phase flips for $\gamma=1$ (black solid lines) and $\gamma=10^3$ (red dash lines). Left pannels are related to $N=6$, right panels to $N=8$. The size of the 1D space $L=1$.}
\label{fstrong2}
\end{figure}

\begin{figure}
\includegraphics[width=1.\columnwidth]{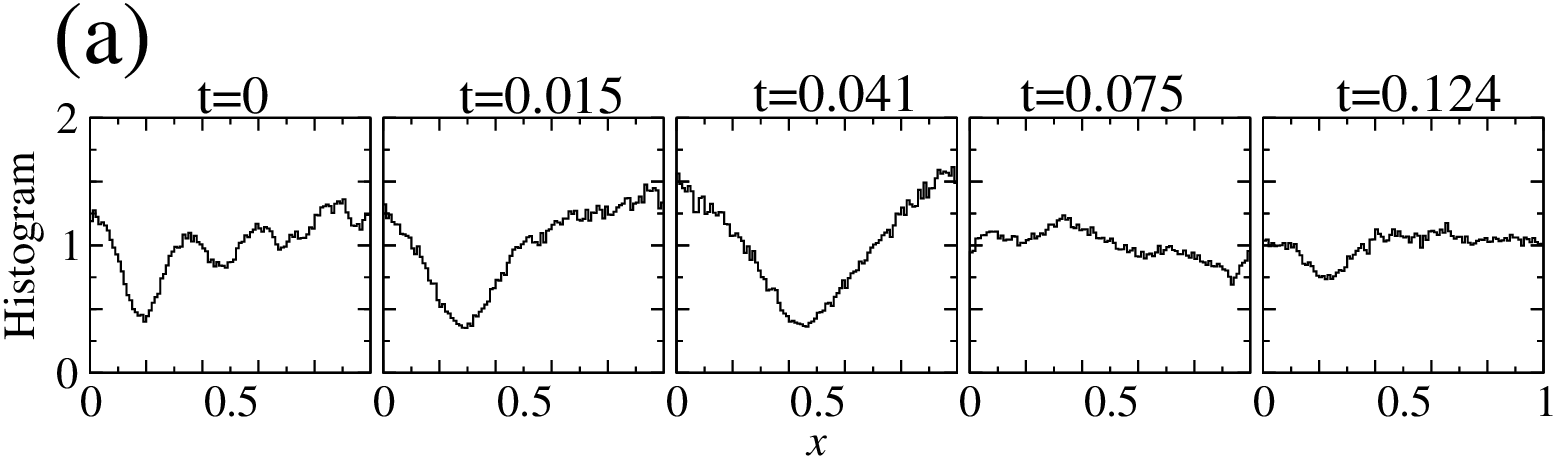}
\includegraphics[width=1.\columnwidth]{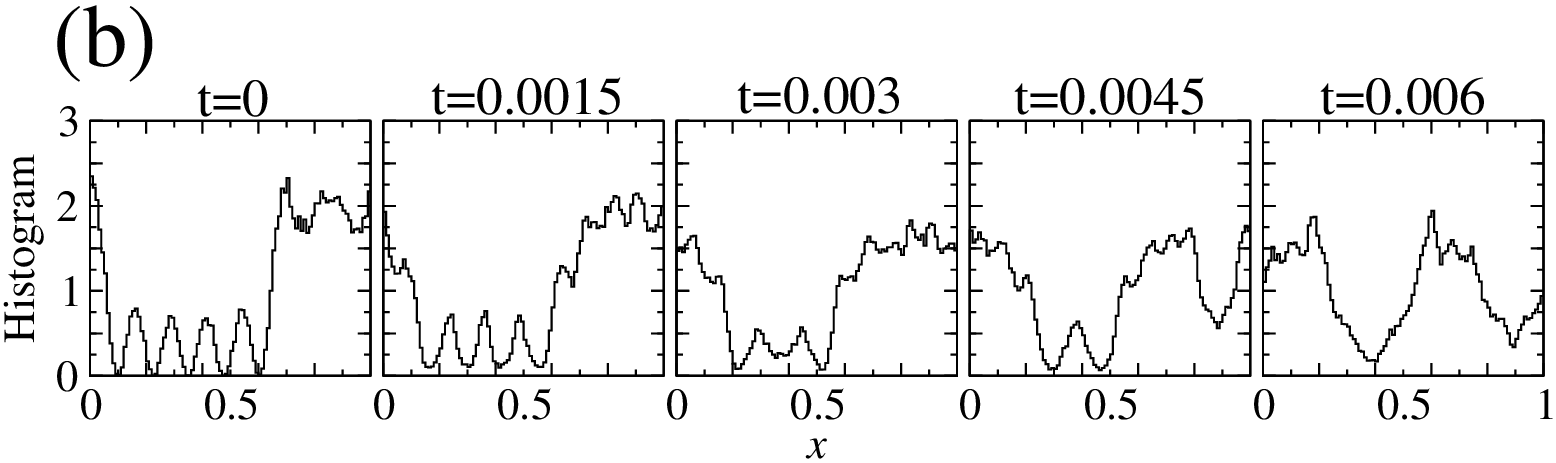}
\caption{{One-hole state} dynamics in a strong interaction regime. Histograms of positions of $N_r=3$ particles obtained in many measurement realizations for different moments of time (indicated in individual panels) after the initial measurement of $N_i=N-N_r$ particles for $\gamma=1$ (a) and $\gamma=\infty$ (b). Initially, $N$-particle system is prepared in the one-hole excited eigenstate with the total momentum per particle $P/N=\pi/L$. In the weak interaction limit such a many-body eigenstate is related to the dark soliton that propagates with the velocity $v=2\pi/L$, cf. Fig.~\ref{fweak_dark_dyn}. For $\gamma=1$, the smearing of the density notch, visible initially in the histogram { at $x\approx 0.25$} [see the first pannel in (a)], occurs on the time scale $1/c=0.125$ that is similar to the period of the dark soliton motion in the mean-field description. For $\gamma=\infty$, five minima are clearly visible in the histogram at $t=0$ [see the first panel in (b)]. These minima correspond to space points where the first $N_r=5$ particles have been measured. At these points it { is not possible} to observe the remaining 3 particles if repulsion between particles is strong. Analysis of the last particle wave-functions shows that the phase flip by $\pi$ is still present and localizes { at different positions (usually at $x>0.6$)} in different measurement realizations. In the course of time evolution, the five minima are blurred very rapidly but the phase flip of the last particle wave-functions is preserved { (see Fig.~\ref{fstrong4}) even at $t=0.75 t_c$ where $t_c\approx0.003$ is a typical lifetime of a generic quantum state, see the text.} The size of the 1D space $L=1$ and $N=8$.}
\label{fstrong3}
\end{figure}

\begin{figure}
\includegraphics[width=0.9\columnwidth]{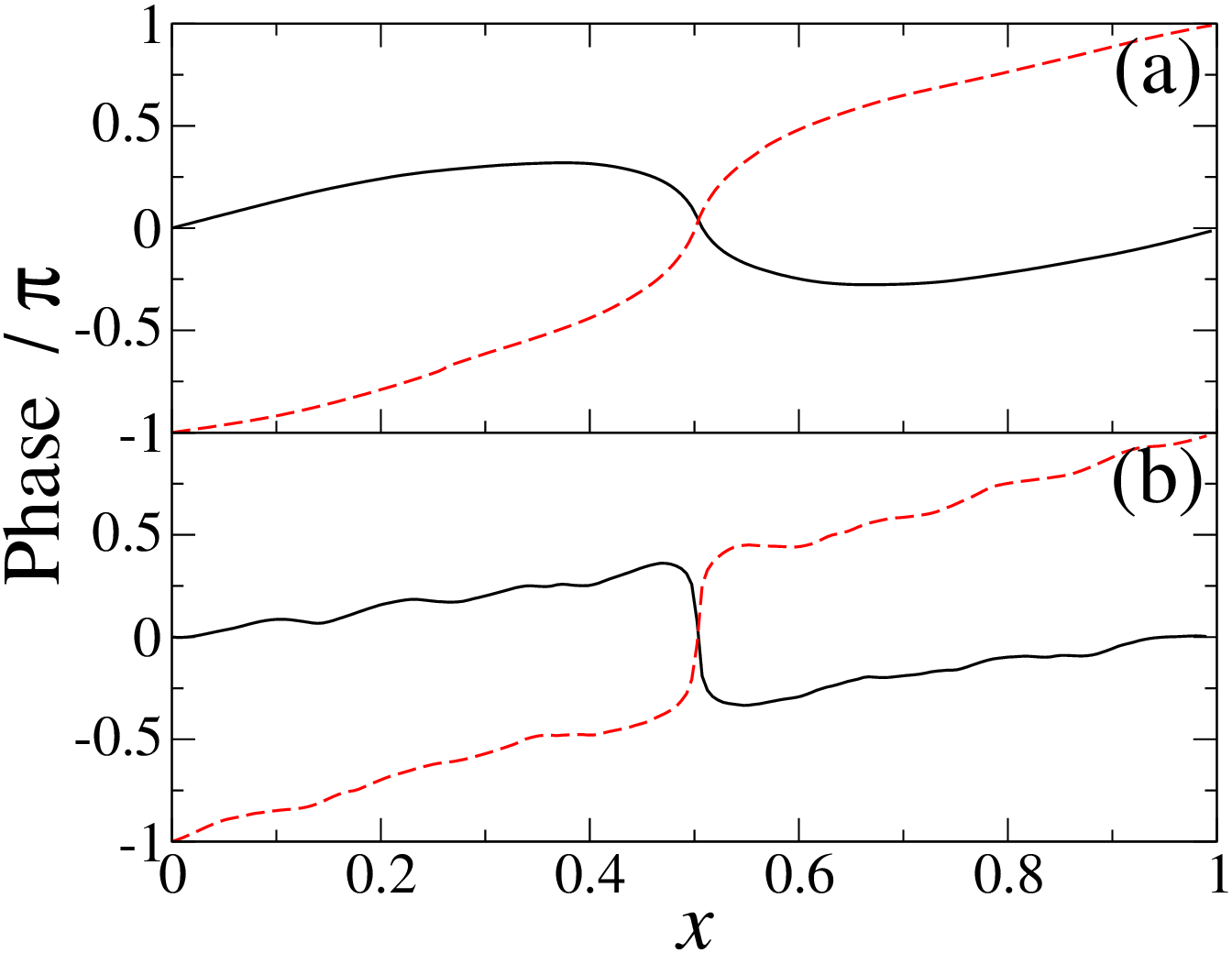}
\caption{(color online) {One-hole state} dynamics in a strong interaction regime. Phase of the last particle wave-functions averaged over many measurement realizations --- the wave-functions with the winding number $J=0$ (black solid lines) and $J=1$ (red dash lines) are averaged independently. The results correspond to one-hole excited eigenstates analyzed in Fig.~\ref{fstrong3}. {Panel} (a) is related to $\gamma=1$ and $t=0.124$ [cf. Fig.~\ref{fstrong3}(a)] while (b) to $\gamma=\infty$ and {$t=0.75t_c=0.00225$} [cf. Fig.~\ref{fstrong3}(b)]. The size of the 1D space $L=1$ and $N=8$.}
\label{fstrong4}
\end{figure}

It is not easy to answer the question if signatures of soliton-like behavior can be observed in the strong interaction limit because the mean-field approach is not valid. In Ref.~\cite{Krutitsky2010} it is shown that the phase imprinting, which is a standard method for dark soliton excitation in experiments, is not able to create dark solitons in ultra-cold atoms in a periodic lattice potential if the system is in the Mott insulator regime. This is due to the lack of phase coherence. On the other hand, analysis of the Lieb-Liniger model indicates that the type~II excitation branch is present regardless how strong the particle interactions are \cite{kanamoto10}. Moreover, superpositions of the one-hole excited eigenstates allows one to create states that reveal density notches in the reduced single particle density and to observe the phase behavior of the matrix elements of the field operator similar as in the weak interaction case \cite{sato12,sato12a,sato16}. In the following we attempt to answer a question if signatures of soliton-like character can emerge in the course of the particle measurements in the system prepared in a one- and two-hole excited eigenstate  for $\gamma\gg 1$.

Let us begin with one-hole excitations. In Fig.~\ref{fstrong1} the average probability density and average phase of the last particle wave-functions, obtained in simulation of the measurement process, are presented for two different one-hole excited eigenstates and for different $\gamma$. The first eigenstate is defined by the exchange $I_{min}\rightarrow(I_F+1)$ in (\ref{gss}) and corresponds to the dark soliton in the weak interaction limit shown in Fig.~\ref{fweak1}(a)-1(b). The other, where $(I_{min}+1)\rightarrow(I_F+1)$, is similar to the gray soliton  analyzed in the weak interaction case, cf. Fig.~\ref{fweak1}(c)-1(f). In each realization we identify a position of the phase flip  and we shift all the wave-functions so that the positions always coincide with $L/2$. The average phase depends very weakly on $\gamma$ and reproduces the behavior observed in the weak interaction limit. The average probability density changes initially with an increase of $\gamma$ but it freezes when $\gamma$ becomes of the order of 100.

For strong repulsive interactions, particles behave like impenetrable bosons and tend to localize with separations $L/N$. The width of the density notches and the period of the density oscillations that are visible in Fig.~\ref{fstrong1}(a)-(b) {for $\gamma=10^3$}, coincide with such a length scale. The oscillations dyes out with an increasing distance from a notch. For $\gamma\rightarrow\infty$, the eigenstates (\ref{ba}) of the Lieb-Liniger model can be approximated by 
\be
\psi_{\{k\}_N}(x_1,\dots,x_N)\propto \prod\limits_{n<m}{\rm sgn}(x_m-x_n){\rm det}\left[e^{ik_jx_s}\right],
\label{bac}
\ee
where $k_j\rightarrow 2\pi j/L$ for odd $N$ \cite{girardeau1960}. The corresponding probability densities are  identical to the probability densities for eigenstates of non-interacting fermions, i.e. $|\psi_{\{k\}_N}(x_1,\dots,x_N)|^2\propto \left|{\rm det}\left[e^{ik_jx_s}\right]\right|^2$. Such a fermionisation is observed in the probability density in the configuration space. Note that the wave-functions (\ref{bac}) remember still the Bose statistics, i.e. they are symmetric under an exchange of any pair of particles.

Results related to the two-hole excitation where $\left\{-I_{min},I_{min}\right\}\rightarrow\left\{-(I_F+1),(I_F+1)\right\}$ in (\ref{gss}), are presented in Fig.~\ref{fstrong2} for two different {even values of} $N$.  We shift the last particle wave-functions so that the position of one of the two observed phase flips coincides with $L/4$. In different realizations of the particle measurements, the two flips appear at different distances one from each other in contrast to the corresponding weak interaction case. 
Histograms of relative distances between the phase flips are depicted Fig.~\ref{fstrong2}. If $\gamma\rightarrow\infty$, the size of the notches is of the order of $L/N$ and the smallest clear separation between the phase flips takes place when only one particle localizes between them. Such cases correspond to the first maximum in the histograms located at $\Delta x$ slightly greater than $L/N$. The next maxima appear at $\Delta x\approx 2L/N$. For $\gamma=1$, the size of the notches is greater than $L/N$ and we can see the remnant of the second notch in the plots of the average densities.

{Analysis of the dynamics is performed in a way similar to the weak interaction case. For $N=8$ we start with the one-hole excited eigenstate corresponding to $P/N=\pi/L$ [cf. Fig.~\ref{fstrong1}(a) and Fig.~\ref{fstrong1}(c)], measure $N_i=5$ particles, evolve the remaining $N_r=3$ particles' reduced state in time and sample the
final probability distribution of positions. The obtained histograms, for different moments of time and for $\gamma=1$ are shown in Fig.~\ref{fstrong3}(a). The density notch, initially visible in the histogram, disappears rapidly, i.e. when $t$ is of the order of $1/c=0.125$. The situation is more complicated when $\gamma\gg 1$. For $\gamma=\infty$, time evolution of a wave-function $\phi(x_1,x_2,x_3)$ of three bosons can be easily obtained by observing that 
\bea
\phi(x_1,x_2,x_3)\prod_{j>s=1}^3 {\rm sgn}(x_j-x_s),
\eea
evolves like a wavefunction of three non-interacting fermions, see (\ref{bac}). The latter evolution can be performed by switching to the momentum space with the help of the Fourier transform. In Fig.~\ref{fstrong3}(b) we show the results for $\gamma=\infty$.
Even at $t=0$, the histogram does not show a clear signature of a soliton-like notch. There are 5 deep minima that correspond to the positions in space  where the first $N_i=5$ particles have been detected --- at these points it is not possible to find remaining 3 particles because of the strong particles repulsion. The phase flip of the last particle wave-functions [very similar like in Fig.~\ref{fstrong1}(c)] localizes randomly at different positions (usually at $x>0.6$). In the course of time evolution the histograms change very quickly, however, the phase flip of the last particle wave-functions can be clearly observed even for {$t=0.75t_c$,} see Fig.~\ref{fstrong4}. The typical lifetime of a generic quantum state $t_c=\pi/2E\approx 0.003$, where $E$ is the averaged energy of the $N_r=3$ particle system with respect to its ground state energy, is related to the so-called speed limit time \cite{sato16}. {In Fig.~\ref{fstrong4}(b) we show average phases related to the winding number $J=0$ and $J=1$. At this moment of time there are also contributions ($\sim24\%$) related to different winding numbers for which position of the phase flip is difficult to identfy.}
The presence of the phase flip is a remnant of the one-hole excitation and a signature that we do not deal with the {Bose} system in the ground state. }

\section{Conclusions}
\label{conclc}

We have considered relation of one- and two-hole excited eigenstates (the so-called type II eigenstates) of the Lieb-Liniger model to dark soliton-like behavior. The eigenstates possess translational symmetry and are not able to reveal solitons. However, the symmetry can be broken in a measurement of positions of particles and the solitons can emerge.

In the weak interaction limit, successive measurements of positions of particles result in a formation of solitons by the remaining particles. It is clearly visible in wave-functions of the last particle, i.e. the wave-functions of the system when all but one particles have been measured. We have also analyzed dynamics of the system prepared initially in a one- or two-hole excited eigenstate. It turns out that when the translational symmetry is broken in a measurement of $N-3$ particles, the remaining three particles follow the corresponding mean-field soliton dynamics for time $t$ smaller than $1/c$. At $t\approx 1/c$ many-body effects, that are beyond the mean-field description, emerge and lead to smearing soliton notches due to a large uncertainty of soliton positions. 

In the strong interaction regime, i.e. when $\gamma\rightarrow\infty$, the Bose system undergoes fermionisation. That is, the particle probability densities become identical to the probabilities of non-interacting fermions. Then, bosons tend to localize with separation $L/N$. Flips of the phase of the last particle wave-functions occur also on the length scale of the order of $L/N$. These phase flips are preserved for a short period of time despite substantial distortion of the probability densities. In the case of two-hole excited eigenstates, the distance between two phase flips is different in different realizations of the measurement process.

Even in the weak interaction limit, the initial many-body eigenstates do not correspond to a Bose-Einstein condensate. However, measurement of positions of particles drives the remaining particles to a state which is the closer to a BEC the more particles were measured. It means that the state of a remaining subsystem approaches a product state with each measurement of particle position. We have considered the cases where in the course of the measurements, the system approaches a BEC with a condensate wave-function corresponding to a single- or multiple-soliton solution of the Gross-Pitaevskii equation. It turns out that such a phenomenon can be observed for a number of particles as small as $N=8$. Numerical simulations, that we perform, are not attainable for a large number of particles. However, we expect the same behavior also for large $N$. Especially, if $N\rightarrow\infty$ but $cN=\mathrm{constant}$, the many-body dynamics is expected to follow
 the mean-field prediction for longer and longer time. That is, the time period when many-body effects turn up should increase like $1/c$. 

Dealing with a small $N$ could be an advantage if one wants to observe many-body effects in soliton dynamics because the time scale needed for these effects to emerge is shorter.

We have considered the emergence of dark solitons from translationally symmetric eigenstates that exist due to the assumed periodic boundary conditions. Such a system can be realized if atoms are loaded to a toroidal trap and prepare in proper excited eigenstates. This is not an easy experiment and its realization requires further research.

\section*{Acknowledgments}
We are gratefull to Krzysztof Giergiel for the help in numerical programming.
The work was performed within the FOCUS action of Faculty of Physics, Astronomy and Applied Computer Science of Jagiellonian University. Support of EU Horizon-2020 QUIC 641122 FET program is also acknowledged. 
Moreover, A.S. acknowledges support in the form of a special scholarship from the Marian Smoluchowski Scientific Consortium "Matter-Energy-Future", from the so-called KNOW funding.
\color{black}

\end{document}